\documentstyle[12pt,epsf]{article}
\newcommand{\sptwo}{1.6}

\newcommand{\doublespace}{\edef\baselinestretch{\sptwo}\Large\normalsize}

\begin{document}

\doublespace

\begin{center}
{\large \bf Single spin measurement in the solid state: a reader for a 
spin qubit}
\end{center}

\medskip

\begin{center}
{\bf S. Bandyopadhyay$\footnote{Corresponding author. E-mail: 
sbandy@vcu.edu}$}\\
{\it Department of Electrical Engineering,
Virginia Commonwealth University,
Richmond, VA 23284 }
\end{center}

\bigskip

We describe a paradigm for measuring a single electron spin in the solid
state. This technique can be used to ``read'' a spin qubit relatively 
non-invasively 
in either a spintronic quantum gate or a spintronic quantum memory. The 
spin reader can be self assembled by  simple electrochemical 
techniques and can be integrated with a quantum gate.

\pagebreak

Measuring single electron spin in a solid is a fundamental problem 
in condensed matter physics. Recently, it has assumed additional importance in 
view of the many spintronic proposals for scalable solid state quantum computers 
that advocate encoding a qubit in a single electron spin  \cite{bandy1, privman, 
loss, 
kane, vrijen, bandy}. In all of these proposals, it will be necessary to measure 
a single electron's spin in order to execute quantum algorithms. This is a 
difficult challenge since unlike charge (which can be
easily measured with electrometers, including the exquisitely sensitive
single electron transistor electrometers), determining a single spin in 
a solid is a formidable challenge.

In quantum computer or memory applications, the requirement is to determine a 
target electron's spin by relatively non-invasive means that are 
``conservative'', meaning that the electron 
should not be lost to a contact (electron reservoir) irretrievably. A basic idea 
might be the following: a target 
spin is coaxed
into tunneling to a region where its wavefunction will overlap with
that of a control spin. It is {\it assumed}  that the
control spin's orientation is known. Since the Pauli principle forbids the 
tunneling event
if the two spins are parallel, the presence or absence of a tunneling
current provides a measurement of the target electron's spin. The difficulty
with this approach is knowing with certainty the orientation of the 
control spin (only a highly localized magnetic field confined to 
within perhaps 10 nm of space around the control spin can orient the
spin deterministically without affecting the target spin. The target qubit has 
to be sufficiently
close (in space) to the control qubit for tunneling to occur and there is 
no known technology to shield a magnetic field over this small distance).
 To our knowledge, there is no report 
of any successful
attempt to demonstrate this reading scheme.

There have been proposals to use a single electron transistor to 
discriminate between a singlet state and a triplet state of a two interacting 
electrons in a solid \cite{australia}. Recently, such a discrimination (using a 
scheme  different from that of ref. \cite{australia}) was demonstrated 
experimentally in a coupled quantum dot system \cite{ono}. However, 
discrimination between singlet and triplet states merely tells us if the 
spins are parallel or antiparallel. It does not tell us which electron has which 
spin, and is therefore not good enough.

In this paper, we describe a paradigm to measure a single electron spin without 
losing that electron to a reservoir irretrievably. The target spin is read via a 
``scout spin'' whose orientation is always known and which interacts directly 
with the measuring device (contacts). The target spin never interacts directly 
with the contacts. Additional advantages are that 
the measuring configuration is completely compatible (and hence can be 
integrated) with an existing model for a 
quantum gate \cite{bandy}.

Consider a penta-layered quantum wire structure as shown in Fig. 1(a). 
The transverse dimension of this wire is $\sim$ 10 nm and the thicknesses
of the semiconducting and insulating layers are also shown. Initially,
the ferromagnetic contacts are {\it not} magnetized, so that the electrons
in the ferromagnetic contacts are {\it not} spin polarized.

\paragraph{Unpolarized spin}: The equilibrium energy band diagram along the 
length of the wire is shown
in Fig. 1(b). We neglect band bending in the semiconductor and insulator
because of resident charges, since such bending may cause small 
quantitative changes, but no qualitative change to the discussion that follows.

The insulating layers are thin enough to be at least translucent to electrons,
but the semiconductor layer is too thick to allow tunneling through it.

If a small potential $V_{SD}$ is applied between the source and drain contacts 
shown in Fig. 1(b), a current will flow only if an electron can jump from the 
source to the lowest subband in the quantum dot, and thence to the drain.
For this discussion, we assume that the temperature is zero (kT = 0) and we also 
neglect all weak virtual processes, so that such a transition is not possible
as long as the lowest subband in the semiconductor dot is {\it above} the Fermi 
level in the source contact.

Let us now pull the lowest subband in the dot to the Fermi level in the source 
contact
by applying a positive potential to the semiconductor, while maintaining 
$V_{SD}$ = 0. This potential can be applied with a wrap-around gate
(in much the same way as in ref. \cite{ono}) or with a remote gate in the 
configuration we will propose later. The gate potential does not affect the 
ferromagnetic metals (because they are ``metals'' which screen the gate 
field), but affects the energy states in the semiconductor (and insulator).
We label the gate potential $V_g$ and the corresponding potential shift that 
it causes in the semiconductor conduction band states is called $V_g'$.
The energy band diagram corresponding to the situation when the lowest
subband in the dot aligns with the Fermi level is shown in Fig. 1(c). For this 
case, the gate potential shift $V_g'$ = $\phi_{ms} + \Delta$. We call this value 
of 
$V_g'$ ($V_g$) = $V_{g1}'$ ($V_{g1}$). When $V_g$ = $V_{g1}$, it becomes 
energetically possible for an electron to jump into the lowest subband 
in the dot from either contact. The dot occupancy now changes from 0 to 1.

Once an electron occupies the dot, it repels a second electron from 
coming into the dot because of Coulomb interaction. For the second 
electron, the lowest available energy state appears to be the level 
shown by the broken line in Fig. 1(c) which is $e/2C$ ($C$ = capacitance 
of the dot) above the lowest subband energy. We have to increase $V_g'$ by an 
additional $e/2C$ to pull the levels down enough so that the broken
line is aligned with the Fermi level in the source as shown in Fig. 1(d). We 
call 
this value of $V_g'$ = $V_{g2}'$ and the corresponding $V_g$ = $V_{g2}$.
Obviously, $V_{g2}'$ = $\phi_{ms} + \Delta + e/2C$. When $V_g$ = $V_{g2}$,
a second electron can enter the dot and occupy it. Pauli Exclusion Principle
dictates that this electron must have its spin {\it anti-parallel} to that of 
the first electron since both electrons are occupying the same lowest subband
of the quantum dot. 

A plot of electron occupancy versus the gate voltage is shown in Fig. 2(a). If
we increase the gate voltage further, beyond $V_{g2}$, ultimately, we will 
pull the second subband level below the Fermi level. Thereafter, more than 
two electrons can occupy the dot, but we shall not explore that region.

\paragraph{Polarized spin}: Now assume that the ferromagnetic contacts 
are taken to their saturation magnetization so that the electrons in them
are spin polarized. Furthermore, assume that the spin polarization is 
100\% (the ferromagnets are essentially half-metallic). Hence, every electron
that enters the dot from the contact has the same spin.

The first electron still enters the dot at $V_g$ = $V_{g1}$, but the 
second electron cannot enter the dot at $V_{g2}$. This is because this
electron has the {\it same} spin as the first, and hence cannot co-exist
in the lowest subband with the first electron because of Pauli Exclusion.
In fact, the second electron can enter the dot only when the gate voltage 
shift $V_g'$ = $\phi_{ms} + \Delta + \Delta'+ e/2C$. We call this value of 
$V_g'$ ($V_g$) = $V_{g3}'$ ($V_{g3}$). The energy band diagram for this 
situation is shown in Fig. 1(e). The charging diagram (or dot occupancy versus 
gate voltage) for {\it spin polarized} electrons is shown in 
Fig. 2(b).

\paragraph{Transport: unpolarized spins} So far, we have assumed that no 
potential is applied across the source and drain contacts ($V_{SD}$ = 0), so 
that no current 
flows. We were merely changing the occupancy of the dot by a gate potential.
Now, let us assume that we apply a small voltage $V_{SD}$ between source 
and drain contacts to induce {\it transport}, so that a non-zero current can 
flow.

The energy band diagrams at different gate voltages $V_{ga}$, $V_{gb}$,
$V_{gc}$, $V_{gd}$ and $V_{ge}$ are shown in Fig. 3 for a fixed value of 
$V_{SD}$. 

When $V_g$ = $V_{ga}$, no current can flow since the intermediate 
state in the quantum dot is not energetically accessible from the source.

When $V_g$ = $V_{gb}$, a current will flow at {\it any} $V_{SD}$ because 
the intermediate state in the dot has become accessible from the source. 
Furthermore,  the drain is also accessible from the intermediate state. This is 
true for any non-zero value of $V_{SD}$.

Now consider the situation in Fig. 3(c) when $V_g$ = $V_{gc}$. The intermediate 
state is 
accessible from the source, but the drain is {\it not accessible} 
from this intermediate state unless $V_{SD} \geq V_t$. If $V_{SD} < V_t$, then 
the Fermi level in the drain is above the subband level in the dot
and therefore electron cannot flow from the dot to the drain. Consequently, 
there 
will be a {\it threshold behavior} in the current-voltage characteristic.
The current voltage characteristic for $V_{gb}$ and $V_{gc}$ are shown in 
Fig. 4(a).

Then, if we increase $V_g$ beyond $V_{gc}$, ultimately the Coulomb repelled 
level (shown by the broken line) will align with the Fermi level in the
source contact (see Fig. 3(d)). We call this value of $V_g$ = $V_{gd}$. Now the 
second 
electron can come in from the source into the dot and escape to the drain
no matter how small $V_{SD}$ is. The first electron is still blocked for $V_{SD} 
< V_t$, but
this does not matter since the second electron (and all following electrons) can 
cause current flow. Thus, the threshold behavior disappears 
at $V_{gd}$. 

It is also obvious that the maximum value of $V_t$ is $e/2C$ and is 
reached just before the gate voltage reaches $V_{gd}$. The dependence of 
the threshold voltage on the gate voltage is shown in Fig. 4(b).

Fig. 4(b) is valid when the two electrons have {\it opposite spins}.  Pauli 
Exclusion  would have prevented the 
second electron from coming into the dot at $V_g$ = $V_{gd}$, if both 
electrons had the same spin.

\paragraph{Transport: spin polarized electrons}. Next, we consider the 
situation when both electrons have the same spin. This corresponds to the 
case when the ferromagnets are magnetized and only one kind of spin can enter 
from the contacts. In this case, electron 1 and electron 2 will have the 
same spin (unless one flips a spin by scattering or because of user 
intervention).

When both electrons have parallel spins, the threshold behavior will not 
disappear until the gate voltage is much larger than $V_{gd}$, and is, in fact, 
equal to $V_{ge}$ as shown in Fig. 3(e). At this gate voltage, the second
subband level aligns with the Fermi level in the source so that a second
electron {\it of the same spin as the first} can enter the dot. This electron 
comes into the second subband since the first subband is not available by virtue 
of the Pauli Exclusion Principle.

The dependence of $V_t$ on $V_g$ {\it for the same spin case} is shown in Fig. 
4(c). It is obvious that the maximum value of $V_t$ in this case is $\Delta'/e + 
e/2C$.

\paragraph{Measurement of single spin}: One can now see how it is possible to 
measure a single electron spin. Our target spin is that of electron 1 and our 
``scout spin'' is that of electron 2. The scout spin comes in from a {\it 
magnetized} ferromagnetic contact and hence its spin is {\it known}. By 
measuring the threshold behavior and discriminating between the cases shown in
Fig. 4(b) and 4(c), we can tell if the target spin is parallel or anti-parallel 
to the scout spin. Hence we can determine the orientation of the target spin. 
Note that the target spin remains trapped in the quantum dot and the scout spin 
goes out to the contact to cause a current. Thus, we can determine the target 
spin somewhat non-invasively without pushing it out to the contact where it
will be lost irretrievably.

\paragraph{Candidate system}: One can synthesize the penta-layered structure
of Fig. 1(a) by sequentially electrodepositing Fe, ZnSe, GaAs, ZnSe and Fe 
within the pores
of a nanoporous alumina film produced by the anodization of aluminum in
sulfuric acid \cite{bandy_nalwa}. We have produced such structures in the past.
Absorption and Raman spectroscopy have independently determined that the 
subband spacing in these dots is about 500 meV \cite{bandy_nanotechnology, 
balandin}. Coulomb blockade experiments in these structures have shown that
the capacitance of the semiconductor layer can be about 0.5 aF, leading 
to a single electron charging voltage $e/2C$ = 160 mV \cite{kouklin}. Therefore,
we can attain the condition $\Delta' > e/2C > kT/e$ at $T$ = 77 K. One can also 
selectively contact a few (about 10) wires by relatively large 
area contacts of 100 $\mu$m $\times$ 100 $\mu$m by exploiting a feature of 
electrochemical synthesis that results in wires of non-uniform height 
\cite{kouklin}. Therefore using 30 $\mu$m $\times$ 30 $\mu$m sized contact
pads (easily made by standard photolithography), one can hope to contact a 
single wire and make the measurements described in this paper. The gate 
potential can be applied by a remote gate that is located far 
away from the source and drain contacts. This structure
is in fact identical to the structure proposed for a universal quantum 
gate in ref. \cite{bandy}. Hence, it can be easily used as a reader of qubit
in that structure.

This work was supported by the National Science Foundation under grant 
ECS-0196554.

\pagebreak


\begin{thebibliography}{10}

\bibitem{bandy1}
S. Bandyopadhyay and V. P. Roychowdhury, {\it Superlattices and 
Microstructures}, {\bf 22}, 411 (1997); S. Bandyopadhyay, A. Balandin,
V. P. Roychowdhury and F. Vatan, {\it ibid}, {\bf 23}, 445 (1998).

\bibitem{privman}
V. Privman, I. D. Wagner and G. Kventsel, {\it Phys. Lett. A}, {\bf 239}, 141 
(1998).

\bibitem{loss}
D. Loss and D. P. DiVincenzo, {\it Phys. Rev. A}, {\bf 57}, 120 (1998).

\bibitem{kane}
B. E. Kane, {\it Nature} (London), {\bf 393}, 133 (1998).

\bibitem{vrijen}
R. Vrijen, et. al., {\it Phys. Rev. A}, {\bf 62}, 12306 (2000).

\bibitem{bandy}
S. Bandyopadhyay, {\it Phys. Rev. B}, {\bf 61}, 13813 (2000).


\bibitem{australia}
B. E. Kane, et. al., {\it Phys. Rev B}, {\bf 61}, 2961 (2000).


\bibitem{ono}
K. Ono, D. G. Austing, Y. Tokura and S. Tarucha, {\it Science}, {\bf 297}, 1313 
(2002).

\bibitem{bandy_nalwa}
S. Bandyopadhyay and A. E. Miller, in {\it Handbook of Advanced Electronic and 
Photonic Materials and Devices}, Ed. H. S. Nalwa, Vol. 6, Chapter 1, (Academic 
Press, San Diego, 2000).

\bibitem{bandy_nanotechnology}
S. Bandyopadhyay, et. al., {\it Nanotechnology}, {\bf 7}, 360 (1996).

\bibitem{balandin}
A. Balandin, K. L. Wang, N. Kouklin and S. Bandyopadhyay, {\it Appl. Phys.
Lett.}, {\bf 76}, 137 (2000).

\bibitem{kouklin}
N. Kouklin, L. Menon and S. Bandyopadhyay, {\it Appl. Phys. Lett.}, {\bf 80},
1649 (2002).
\end{thebibliography}

\pagebreak

\noindent {\bf Fig. 1}: (a) A penta-layered quantum wire structure of transverse 
dimension $\sim$ 10 nm. The  thicknesses of the constituent layers are shown. FM 
= ferromagnet, I = insulator and SC = semiconductor. (b) Equilibrium energy band 
diagram along the length of the wire. $\phi_{ms}$ is the metal semiconductor 
work function difference, $\Delta$ is the quantization energy of the first 
subband and $\Delta'$ is the difference between the quantization energies of 
the second and first subband. (c) Energy band diagram when a gate potential 
$V_{g1}$
pulls the semiconductor (and insulator) conduction band down to make the 
lowest subband level in the semiconductor quantum dot align with the 
Fermi level. A single electron can now occupy the dot. (d) Energy band diagram 
when the gate potential $V_{g2}$ pulls the quantum dot levels down by an 
additional 
$e/2C$ ($C$ = dot capacitance). Now two electrons can occupy the dot {\it if 
they have opposite spins}. (e) Energy band diagram when the second subband is 
pulled down flush with the Fermi level by gate potential $V_{g3}$. Now two 
electrons of the {\it same spin} can occupy the dot.

\bigskip

\noindent {\bf Fig. 2}: Charging diagram (dot occupancy versus gate voltage) for 
(a) spin-unpolarized electrons and (b) spin-polarized electrons.

\bigskip

\noindent {\bf Fig. 3}: Electron energy band diagram at different gate voltages 
(a) $V_g$ = $V_{ga}$ when no current can flow since the intermediate dot state 
is not
accessible from the source. (b) $V_g$ = $V_{gb}$ when the first subband lines up 
with the Fermi level in the source. Now the intermediate state is accessible 
from the source and the drain is accessible from this intermediate state. Hence, 
a current can flow at 
{\it any} value of $V_{SD}$. (c) $V_g$ = $V_{gc}$ when the first subband level
dips below the Fermi level at the source. At this point the intermediate state 
is accessible from the source, but the drain is not accessible from this 
intermediate state unless $V_{SD}$ exceeds a certain threshold value. (d) $V_g$ 
= $V_{gd}$ when the Coulomb repelled level (broken line) lines up with the Fermi 
level in the source. In this case, current can flow when $V_{SD} > e/2C$ so that 
the threshold voltage $V_t$ = $e/2C$. This is the maximum value of the threshold 
voltage. If we increase the gate voltage further, the second electron can come 
in from the source and conduct current at any $V_{SD}$ thereby collapsing the 
threshold behavior. This will happen only if the two electrons have opposite 
spins so that Pauli blockade is not operative. If the second electron has the 
same spin as the first, it cannot come into the dot because of Pauli exclusion 
and hence the threshold behavior will continue and not collapse. (e) $V_g$ = 
$V_{ge}$ when the second subband level lines up with the Fermi level in the 
source. Now the second electron can enter the dot even if it has the same spin
as the first because it is {\it not} occupying the same subband. At this point,
the threshold behavior will collapse even if the electrons have parallel spins.

\bigskip

\noindent {\bf Fig. 4}:(a) Current voltage characteristic showing a threshold 
behavior. 
Threshold voltage $V_t$ versus gate voltage $V_g$ when (b) the electrons have 
anti-parallel spins, and (c) the electrons have parallel spins.

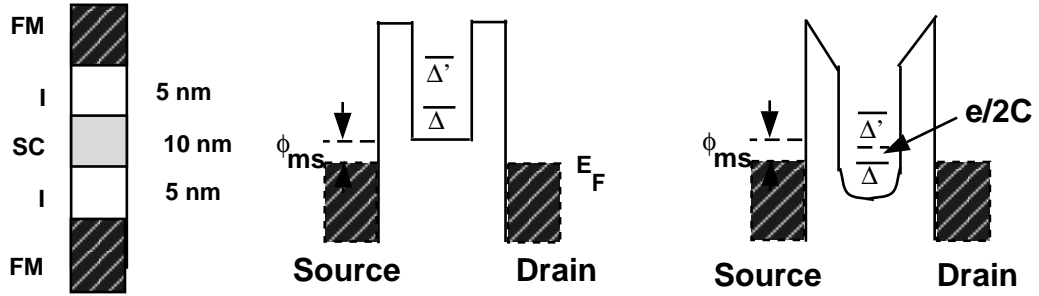
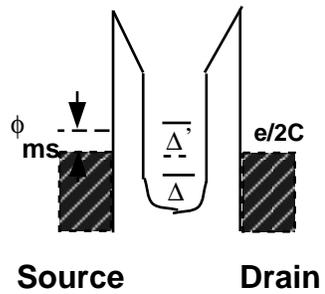
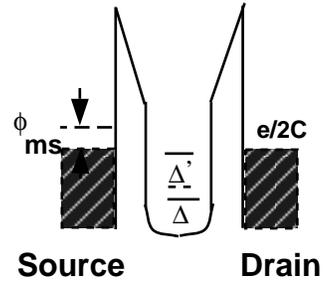
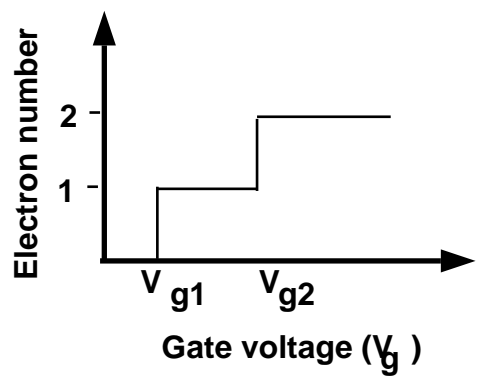
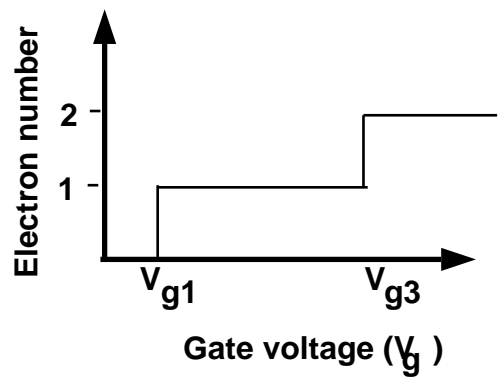
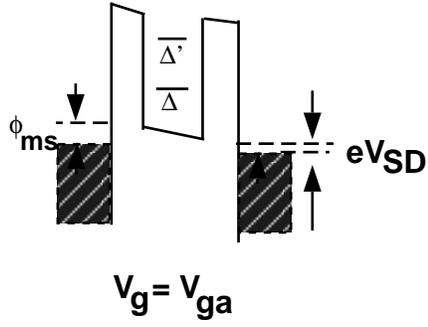
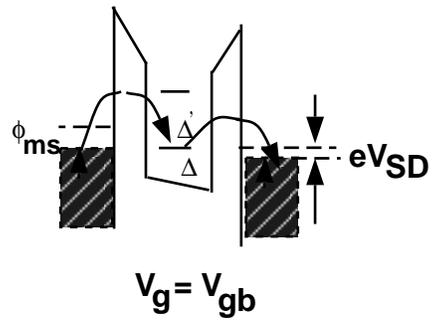
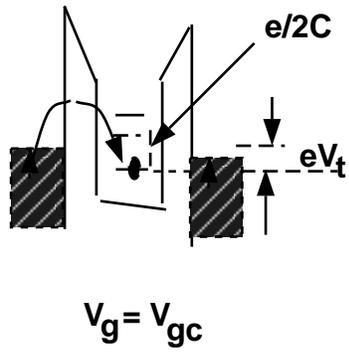
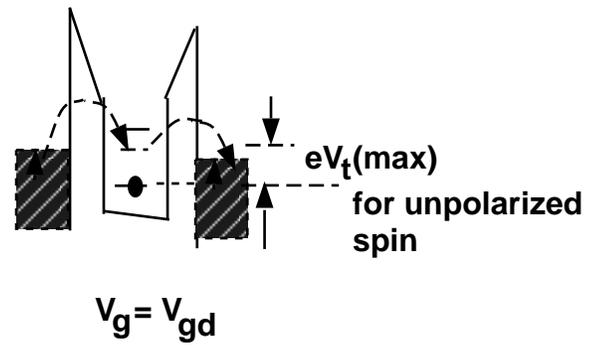
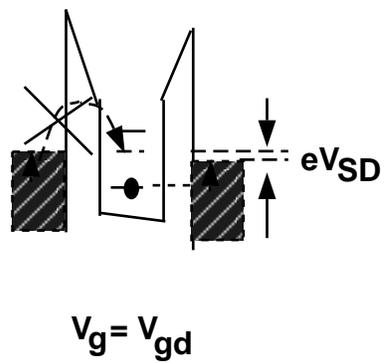
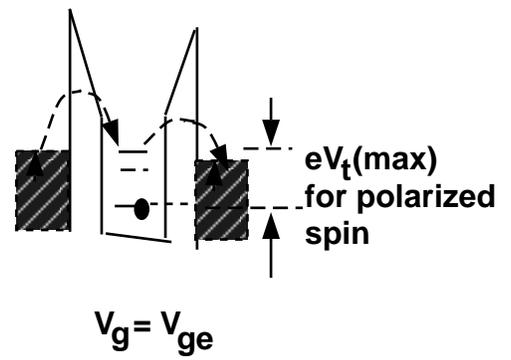
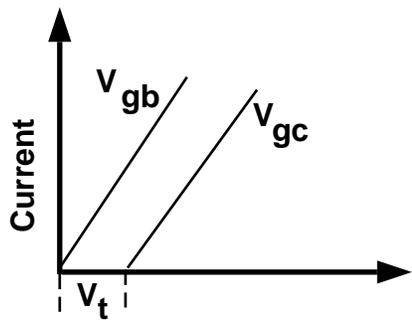
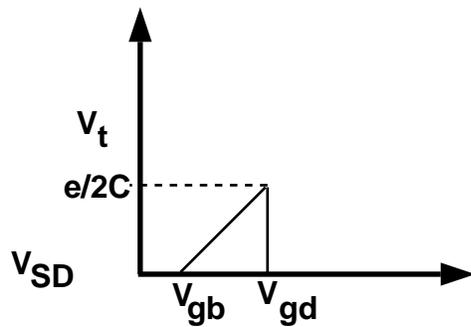
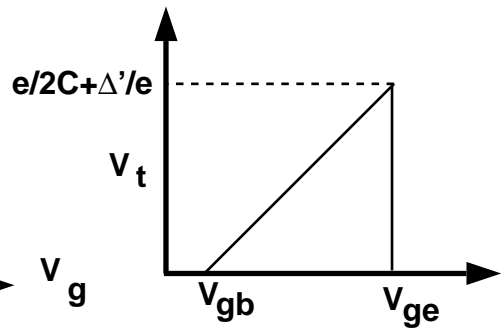
\end{document}